\pdfoutput=1
\documentclass[fleqn,usenatbib]{mnras}
\usepackage[T1]{fontenc}
\usepackage{ae,aecompl}
\usepackage{graphicx}
\usepackage{amsmath}
\usepackage{amssymb}
\usepackage{txfonts}

\newcommand{\msun}{\textnormal{M}_\odot}
\newcommand{\rsun}{\textnormal{R}_\odot}

\defcitealias{RH17}{H17}

\title[Origin of BH in iPTF13bvn]{The Origin of the Possible Massive Black Hole in the Progenitor System of iPTF13bvn}

\author[R. Hirai]{
Ryosuke Hirai$^{1}$\thanks{E-mail: hirai@heap.phys.waseda.ac.jp}
\\
$^{1}$Faculty of Science and Engineering, Waseda University, 3-4-1, Okubo, Shinjuku, Tokyo 169-8555, Japan\\
}

\date{Accepted XXX. Received YYY; in original form ZZZ}

\pubyear{2017}

\begin{document}
\label{firstpage}
\pagerange{\pageref{firstpage}--\pageref{lastpage}}
\maketitle

\begin{abstract}
 This letter complements a formation scenario of the progenitor of the supernova iPTF13bvn proposed in \citet{RH17}. Although the scenario was successful in reproducing various observational features of the explosion and pre-explosion photometry by assuming that the progenitor had a relatively large black hole companion, it lacked an explanation for the origin of the black hole itself. We now explore the possible evolutionary paths towards this binary with a relatively large black hole companion. We found that the black hole was probably produced by a very massive star that experienced common envelope evolution. According to our \texttt{MESA} stellar models, the primary mass should have been $\gtrsim70\msun$ to reproduce the required remnant mass and final separation. This indicates that iPTF13bvn was likely a rare case and normal type Ib supernovae originate from different paths.
\end{abstract}

\begin{keywords}
stars: massive -- binaries: close -- supernovae: individual: iPTF13bvn
\end{keywords}

\section{Introduction}
 Core-collapse supernova (SN) explosions are considered to be the final fate of massive stars with masses $\gtrsim8\msun$. The majority of these massive stars are known to be in close binary systems \citep[]{chi12,san14}. Thus the appearance and features of SNe will be strongly affected by binary interactions. It is considered to be especially important for understanding the origin of the so-called stripped-envelope SNe. This is a certain class of SNe that have lost a significant amount of their hydrogen envelope prior to the explosion, usually classified into type Ib, Ic or IIb by their spectra. Although it is still not settled whether the loss of the envelope is due to binary interactions or strong stellar winds, the small observed ejecta mass strongly supports the binary scenario \citep[]{lym16}.

iPTF13bvn is a type Ib SN that has been extensively studied due to the fact that it is the only type Ib SN known to have a corresponding pre-SN image detection. Soon after its discovery by \citet{cao13}, some single Wolf-Rayet star progenitor models were proposed that were consistent with the early pre-SN image \citep[]{gro13}. However, the following fast decline of the light curve suggested an ejecta mass of $\sim2\msun$, that could not be explained by these models but instead favoured binary progenitor models \citep[]{ber14,fre14,sri14,eld15,kun15}. Later observations of the SN site were carried out by \citet{fol16}, and revealed that the observed flux has declined below the pre-SN flux. This means that the pre-SN source was indeed the progenitor itself, and not a nearby star that was aligned with the SN by coincidence. Even more, the upper limit placed by the new observations were strong enough to conclude that there was no bright companion to the progenitor, refuting most existing binary evolution models at the time \citep[]{ber14,RH15}.

It was shown by \citet{eld16} that according to their huge grid of binary evolution calculations, scenarios involving common envelope (CE) episodes can produce similar progenitors but with relatively dark companions. However, \citet[][hereafter \citetalias{RH17}]{RH17} derived a lower limit to the progenitor radius from the pre-SN photometry and argued that CE scenarios cannot reproduce the large radius\footnote{CE models involving jets launched from the inspiralling companion may have enough energy to eject the envelope \citep[]{pap15,sok15}. However, such processes are still poorly understood and are difficult to derive concrete conclusions.}. An alternative scenario was demonstrated, where the companion is a large black hole (BH). If the BH mass is larger than a certain fraction of the primary mass, the system will undergo stable mass transfer (MT) and will strip off a large amount of the envelope. It will leave an almost naked helium star by the time of explosion, and the companion will not show up in the remnant after the explosion. In this way the progenitor system can satisfy all constraints placed by the pre-SN image, light curve and late time photometry. However, they only demonstrated a sample evolutionary track of a binary with a BH component and did not thoroughly discuss the origin of the BH, although it is not so obvious.

In this letter, we provide support to the scenario in \citetalias{RH17} by showing possible evolutionary paths towards the massive BH. We aim to construct a self-consistent evolutionary scenario that extends from the formation of the binary up to the SN explosion. This letter is structured as follows. In Section \ref{sec:previous} we will briefly review the BH companion scenario proposed in \citetalias{RH17}. Then we discuss the origin of the BH companion in Section \ref{sec:origin} and summarize our results in Section \ref{sec:conclusion}.

\section{Black Hole Companion Scenario}\label{sec:previous}
Here we briefly summarize the BH companion scenario proposed in \citetalias{RH17}. In their demonstration, they start off the evolution with a binary consisted of a 16 $\msun$ star with a 15 $\msun$ BH companion in an 8~d orbit. As the primary star evolves and starts burning helium at the core, it eventually fills its Roche lobe and transfers matter to the BH. This is called case B MT, and will continue until the star detaches from its Roche lobe with only a tiny portion of hydrogen left. It will then carry on its evolution up to core-collapse without filling its Roche lobe again. The BH will accrete a fraction of the matter but will also eject a part of it out of the system carrying away angular momentum. It is not known how much matter can be accreted on to the BH, but in \citetalias{RH17} they limit the accretion rate to the Eddington limit. Many researches show that super-Eddington accretion is possible \citep{ohs05,jia14,sad16}, but even if we assume higher accretion rates, it does not strongly affect the evolution of the primary \citep{dev12,ben13}. Furthermore, they checked that the evolution of the primary is almost totally insensitive to the BH mass as long as the mass ratio is $q_\textrm{BH}\equiv m_\textrm{BH}/m_\textrm{p}\gtrsim0.8$, where $m_\textrm{BH}$ is the BH mass and $m_\textrm{p}$ is the primary star mass. The overall evolution did not change even with BH masses $\sim100\msun$.

The sample evolutionary track in \citetalias{RH17} was successful in reproducing the pre-SN photometry, small ejecta mass and satisfying the post-SN photometry limits. However, there is a range of initial parameters that can follow similar tracks. In \citetalias{RH17} they roughly estimate that the primary mass should have been in the range $\sim14$--$17\msun$ to produce a $\sim3.5$-$\msun$ helium core, and the initial period $\sim4$--20~d to initiate case B MT although case A scenarios can also be possible.

\begin{figure}
 \centering
 \includegraphics[width=\linewidth]{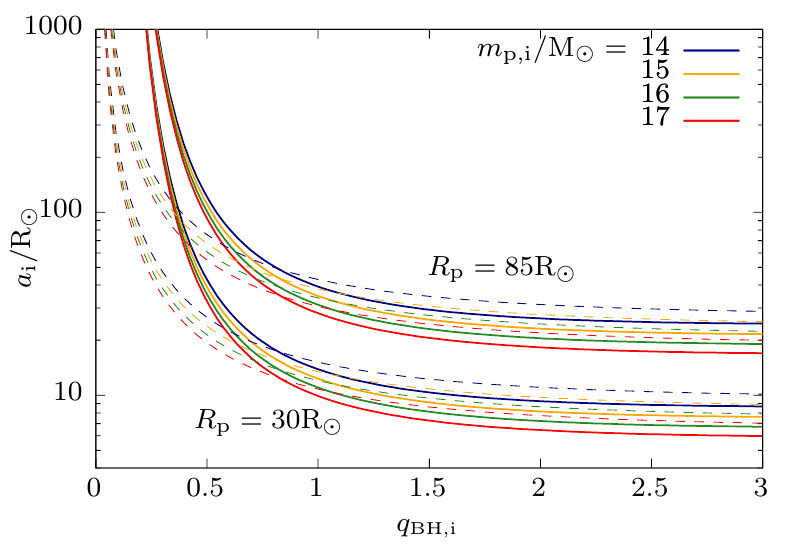}
 \caption{Initial separation of binaries that end with a primary Roche lobe radius $R_\textrm{p}=30\rsun$ (lower set) and $R_\textrm{p}=85\rsun$ (upper set). Line colours indicate the initial mass of the progenitor. Lines are drawn assuming conservative MT (dashed) and fully non-conservative MT (solid).\label{fig:sep}}
\end{figure}

To provide stricter constraints on the binary configuration, we will re-evaluate the limit on the initial separation more properly. The binary separation changes when mass is transferred between the binary components or any mass is lost from the system. If we assume that all the matter transferred to the BH is expelled from the system (fully non-conservative MT) with the specific angular momentum of the BH, we can calculate the evolution of the orbit analytically and relate the initial and final separations as
\begin{align}
 a_\textrm{i}=a_\textrm{f} \frac{m_\textrm{f}}{m_\textrm{i}}\left(\frac{m_\textrm{p,f}}{m_\textrm{p,i}}\right)^2\exp\left(2 \frac{m_\textrm{p,i}-m_\textrm{p,f}}{m_\textrm{BH}}\right) \label{eq:inisep}
\end{align}
where $a$ is the separation, $m=m_\textrm{p}+m_\textrm{BH}$ is the total mass and the subscripts i and f denote the initial and final states \citep[e.g.][]{pos14}. On the other hand if we assume that all of the transferred matter can be retained by the BH (fully conservative MT), the relation between the initial and final separations can be expressed as 
\begin{align}
 a_\textrm{i}=a_\textrm{f}\left(\frac{m_\textrm{BH,f}m_\textrm{p,f}}{m_\textrm{BH,i}m_\textrm{p,i}}\right)^2
\end{align}
Among these values, the final and initial progenitor masses are known to be $m_\textrm{p,f}\sim3.5\pm0.5\msun$ from light curve modelling \citep[]{ber14,fre14,sri14} and $14\lesssim m_\textrm{p,i}\lesssim17$ from our binary evolution model respectively \citepalias{RH17}. Here we will take $m_\textrm{p,f}=3.5\msun$ for clarity. Because the final Roche lobe radius should be larger than the progenitor, we can also place a lower limit to the final separation $a_\textrm{f}$ by using the Eggleton relation \citep[]{egg83}
\begin{align}
 a_\textrm{f}\geq R_\textrm{p,f}\frac{0.6q_\textrm{BH,i}^{-2/3}+\ln(1+q_\textrm{BH,i}^{-1/3})}{0.49q_\textrm{BH,i}^{-2/3}} \label{eq:eggleton}
\end{align}
where $R_\textrm{p,f}$ is the progenitor radius at the end of the MT. These equations give us a lower limit on the initial separation as a function of the mass ratio, which we plot in Figure~\ref{fig:sep}. For our fiducial model we use $R_\textrm{p}=85\rsun$ which is the progenitor radius just before it detaches from the Roche lobe (i.e. when the MT ends) in the sample evolutionary track shown in \citetalias{RH17}. We also show the lowest possible limit using $R_\textrm{p,f}=30\rsun$ which is the minimum pre-SN progenitor radius. The value of $R_\textrm{p,f}$ is roughly related with the mass of the remaining envelope, which in turn determines the pre-SN radius. However, the mass of the remaining envelope can change due to wind mass-loss in the Wolf-Rayet phase so it is hard to firmly constrain the value of $R_\textrm{p,f}$. Here we just show the representative models but we can regard these as lower limits because the evolution calculations in \citetalias{RH17} did not include strong wind mass-loss for the Wolf-Rayet phase. Note that $a_\textrm{i}$ is directly proportional to $R_\textrm{p,f}$. By comparing the solid and dashed lines, it is evident that the fraction of matter accreted on to the BH is not so important at least at $q_\textrm{BH,i}>0.8$. It can be seen that the initial separation should be at least $a_\textrm{i}\gtrsim17\rsun$ which translates to an orbital period of $P_\textrm{i}\gtrsim1$~d, or even larger depending on the primary and BH masses (e.g. $a_\textrm{i}\gtrsim50\rsun$ for $m_2=14\msun, q_\textrm{BH,i}=0.8$). This initial separation is large enough to rule out case A MT cases. If the initial separation is too large, the primary will reach its Roche lobe at more evolved states, where convective envelopes may develop. MT from convective envelopes are known to be unstable and leads to CE phases, so the lower $q_\textrm{BH,i}$ models are not appropriate. Here we will only consider the $q_\textrm{BH,i}\gtrsim0.8$ models to avoid further complications.

\section{Origin of the Black Hole}\label{sec:origin}
The model in the previous section seems to be a promising candidate to explain the formation of the progenitor of iPTF13bvn. However, this scenario is based on the assumption that the primary star had a massive BH companion, which originally should have been a massive star too. Namely, the model only discusses the second stage of the evolution of the binary, but there should have been a first stage of evolution where an initially more massive star collapsed to a BH after interacting with the less massive companion. Here we will discuss this first stage in order to complement this scenario in a self-consistent way. In the following we will denote the progenitor of the BH as the \textit{primary} with a mass $m_1$, and the progenitor of the SN as the \textit{secondary} with a mass $m_2$. We will call the second stage of the evolution as the X-ray binary (XRB) phase, because it has a configuration similar to X-ray binaries.

\subsection{BH formation in general}

BHs are considered to be descendants of massive stars that experience core-collapse at the end of their evolution. However, the mass range in which stars can collapse to BHs are highly uncertain. Stellar evolution theory suggests that stars more massive than $\gtrsim25\msun$ can produce BHs after a weak core-collapse SN explosion by fallback accretion on to the central neutron star \citep[]{fry99,heg03}. Only the low metallicity stars will be able to directly collapse to BHs without explosions. The massive stars with solar metallicity will experience a large amount of mass loss by stellar winds, and will have lost most of or its entire hydrogen envelope by the time of collapse, producing type Ib or Ic SN explosions. However, the average observed ejecta mass of stripped-envelope SNe is quite small ($\sim2\msun$) which contradicts with the massive single-star models \citep[]{lym16}. It may mean that a larger fraction of these massive stars collapse to BHs than anticipated even if they have high metallicities.

According to these theoretical predictions, stars less massive than $\lesssim25\msun$ are able to produce successful type IIP SN explosions. However, the maximum mass of the known progenitors is $\lesssim16.5\pm{1.5}\msun$, showing a deficit of high-mass progenitors \citep[e.g.][]{sma09}. Although it may be explained by observational biases such as strong dust extinction around red supergiants \citep[]{bea16}, the most natural interpretation will be to assume that stars with masses $18\lesssim M/\msun\lesssim25$ cause failed SNe and produce BH remnants \citep[]{hor14,koc14,koc15}. The disappearance of a $\sim25\msun$ red supergiant has recently been reported, which may be strong support for this hypothesis \citep[]{ada16a,ada16b}. The details of failed SNe are also uncertain, but it is predicted that the brief neutrino emission will lead to a sudden loss of gravitational mass, and cause the loose hydrogen envelope to become unbound \citep[]{nad80,lov13}.

To sum up, the recent observational facts suggest that most stars larger than $\gtrsim18\msun$ are likely to collapse to BHs. This roughly corresponds to the mass range in which core-collapse SN progenitor models have structures that are more difficult to explode \citep[]{oco11,ugl12,suk16,ert16}. For the failed SN cases ($18\lesssim M/\msun\lesssim25$), the hydrogen envelope can be lost due to the sudden loss of gravitational mass by neutrino emissions. Thus the remnant mass will be roughly $90\%$ of the pre-collapse helium core mass. For the directly collapsing cases ($\gtrsim25\msun$), the remnant mass will be equivalent to its pre-collapse stellar mass, which strongly depends on the highly uncertain wind mass loss rate \citep[]{fry12}.

In order to produce the large BH ($m_\textrm{BH,i}=m_\textrm{p}q_\textrm{BH,i}\gtrsim11.2\msun$) for the following XRB phase in our scenario, failed SN channels are insufficient so the progenitor of the BH should have been at least $\gtrsim25\msun$ and experienced direct collapse.

\subsection{Stable Mass Transfer Paths}

Here we will discuss the formation of the possible BH in the progenitor system of iPTF13bvn. Because our aim is to form a BH-massive star binary with an orbital separation close enough to undergo case B MT, the first stage of evolution should also have involved binary interactions. There are in general two channels of binary interactions: stable MT and CE evolution. In order to have stable MT, the initial binary mass ratio needs to be close to unity. The critical mass ratio is usually assumed to be $q_1=m_1/m_2\sim3.5$ \citep[]{web85,hje89,iva04,ge10} where a delayed dynamical instability may form. But even if the MT is stable, the binary may enter a de facto CE phase if the secondary star cannot accumulate the transferred matter \citep[]{iva13}. This will mainly happen when the accretor is not a compact object, and cannot eject the accreted matter from the system. In order to retain all the transferred matter, the thermal time-scale of the accretor needs to be close to the MT time-scale which is roughly determined by the thermal time-scale of the donor.  For the system in concern now, the primary star needs to be at least $\gtrsim25\msun$ to produce the large BH, whereas the companion needs to be $\sim14$--$17\msun$ at the end of MT (cf. Section \ref{sec:previous}). We are not sure of the initial mass ratio but it is likely to be $q_1\gtrsim2$, which makes a considerable difference in the thermal time-scales of the stars. Thus we consider that the stable MT channel is unlikely for this system.

\subsection{Common Envelope Paths}

When the mass ratio is relatively large or the initial separation is large so that the donor envelope becomes convective, the binary is likely to experience unstable MT which leads to CE phases. A CE phase is where one of the stars in a binary engulfs its companion and the engulfed star plunges in towards the centre due to dynamical friction with the envelope matter. If there is enough energy in the orbit to unbind the envelope, the system will be left with the core of the primary and the companion in a tighter orbit. Otherwise the stars will simply merge into a single object.

For the case here, we need to consider the outcome of a CE phase with an initially very large ($m_1\gtrsim25\msun$) primary star with a $m_2\sim$14--17$\msun$ companion. After the CE phase, the system will be left with the naked helium core of the primary orbiting the secondary in a tight orbit. Then the naked helium star will evolve as a Wolf-Rayet star, which will expel a large fraction of its mass by strong stellar winds. The wind will take away angular momentum and widen the orbit until the primary collapses into a BH. At this point the primary should have a mass $m_1\gtrsim0.8m_2$ and the binary separation larger than $a_\textrm{i}$ given in Figure \ref{fig:sep}.

In order to test the plausibility of this scenario, we carry out stellar evolution calculations using a public code \texttt{MESA} \citep[version 8845;][]{MESA1,MESA2,MESA3}. For convection we use the mixing length theory with the Ledoux criterion and a mixing length 1.6. We limit the mass of mass shells to be $<0.01\msun$ which is the resolution recommended to properly evaluate helium core masses with \texttt{MESA} \citep[]{far16}. There are two points that need to be investigated - the remnant mass and pre-SN separation. To evaluate the remnant mass, we follow the evolution of massive stars that experience CE phases by taking the same procedures as in \citet{iva11} and \citetalias{RH17}. We first follow the evolution of single stars up to the point where the stellar radius exceeds $600\rsun$\footnote{The critical radius here is not important because the core mass does not grow during the expansion \citepalias{RH17}.}, where we assume that the binary enters a CE phase. At this point we record the mass coordinate $m_\textrm{c}$ of the \textit{maximum compression point} \citep[]{iva11} and apply a mass loss rate of $\dot{M}\sim-1\msun$yr$^{-1}$ until the mass reaches this value. Then we leave the almost naked helium star to evolve until it starts burning silicon at the centre.

\begin{figure}
 \centering
 \includegraphics[width=\linewidth]{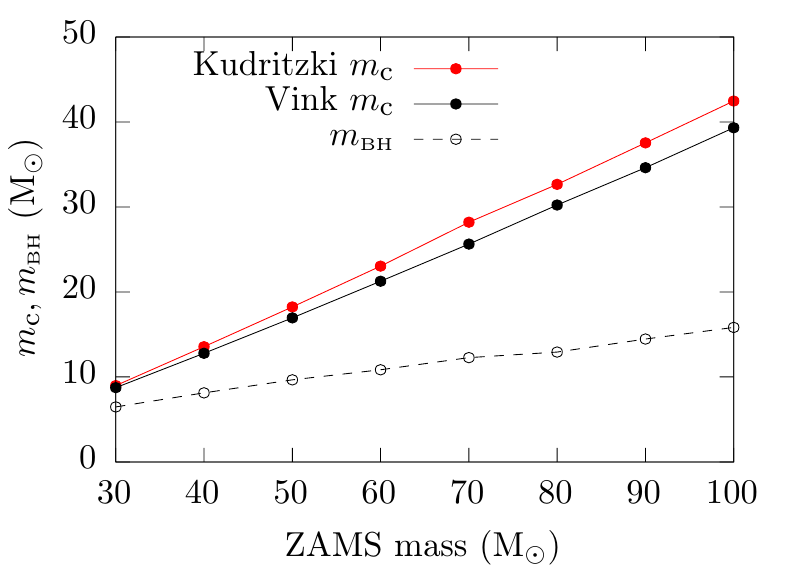}
 \caption{Core masses at the onset of the CE phase in our \texttt{MESA} runs. Colours of the lines indicate which wind mass loss scheme was used. The dashed line shows the final BH mass obtained after a WR phase in the Vink model.\label{fig:remmass}}
\end{figure}

The BH mass will be equivalent to the remaining stellar mass at the final stage, which is determined by the wind mass loss rate. In our models we use two different prescriptions for main-sequence mass loss: one by \citet[][hereafter Kudritzki scheme]{kud89} and one by \citet[][hereafter Vink scheme]{vin01}. Figure \ref{fig:remmass} shows the pre-CE helium core mass $m_\textrm{c}$ obtained in our \texttt{MESA} calculations. Both schemes give similar core masses. However, after the CE phase, the star will enter a Wolf-Rayet phase where its mass-loss rate is highly uncertain. Many stellar evolution calculations adopt the empirical formula by \citet{nug00}, but this is a rough estimate since the observed values still have a wide scatter. The dotted line shows the final remnant mass obtained using the empirical formula on our Vink models assuming solar metallicity\footnote{The progenitor was likely to have been formed in a solar-metallicity environment \citep[]{fre16}.}. It can be seen that roughly half of its mass can be lost in the Wolf-Rayet phase, and only the extremely massive models (ZAMS mass $\gtrsim70\msun$) have remnant masses that satisfy the requirement in our BH companion scenario. However, it should be emphasized that this is just a sample model and the BH mass can be larger depending on the unknown wind mass loss rate.

We estimate the final separation of the binary with the so-called ``energy formalism'' \citep[]{web84,ibe84},
\begin{align}
 E_\textrm{env}=\alpha_\textrm{CE}\left(-\frac{Gm_\textrm{c}m_2}{2a_\textrm{f}}+\frac{Gm_1m_2}{2a_\textrm{i}}\right)
\end{align}
where $E_\textrm{env}$ is the binding energy of the primary, $G$ is the gravitational constant, $m_1, m_2$ are the masses of the primary and secondary respectively and $a_\textrm{i}, a_\textrm{f}$ are the binary separations before and after the CE phase. $\alpha_\textrm{CE}$ is an efficiency parameter of the energy conversion which is usually taken as unity. The second term in the parentheses is usually negligible compared to the first term because the post-CE separation is much shorter. Therefore we can estimate the post-CE separation as
\begin{align}\label{eq:sep}
 a_\textrm{post-CE}\sim\frac{\alpha_\textrm{CE}Gm_\textrm{c}m_2}{2E_\textrm{env}}
\end{align}
where $m_\textrm{c}$ and $E_\textrm{env}$ can be calculated from our evolution simulations. The binding energy of the envelope is calculated by comparing the total binding energy of the star before and after the CE event to take into account the relaxation of the core after the mass ejection \citep[]{ge10}.
\begin{align} \label{eq:Eenv2}
 E_\textrm{env}&=E_\textrm{bind,i}-E_\textrm{bind,f}\nonumber\\
&=-\int_0^{m_1}\left(-\frac{Gm}{r}+\epsilon\right)dm+\int_0^{m_\textrm{c}}\left(-\frac{Gm}{r}+\epsilon\right)dm
\end{align}
Here $\epsilon$ is the specific internal energy, $r$ is the radius and the integrations are taken over the whole star before (first term) and after (second term) the CE phase. After the CE phase, the naked helium star will emit strong stellar winds which will widen the separation. If we assume that the wind takes away angular momentum proportional to the specific angular momentum of the star, the final separation can be calculated as 
\begin{align}\label{eq:finsep}
a_\textrm{f}=a_\textrm{post-CE}\frac{m_c+m_2}{m_\textrm{BH}+m_2}
\end{align}
Because the final BH mass is highly uncertain, we take the lower limit $m_\textrm{BH}=0.8m_2$ to derive an upper limit to the final separation.

\begin{figure}
 \centering
 \includegraphics[width=\linewidth]{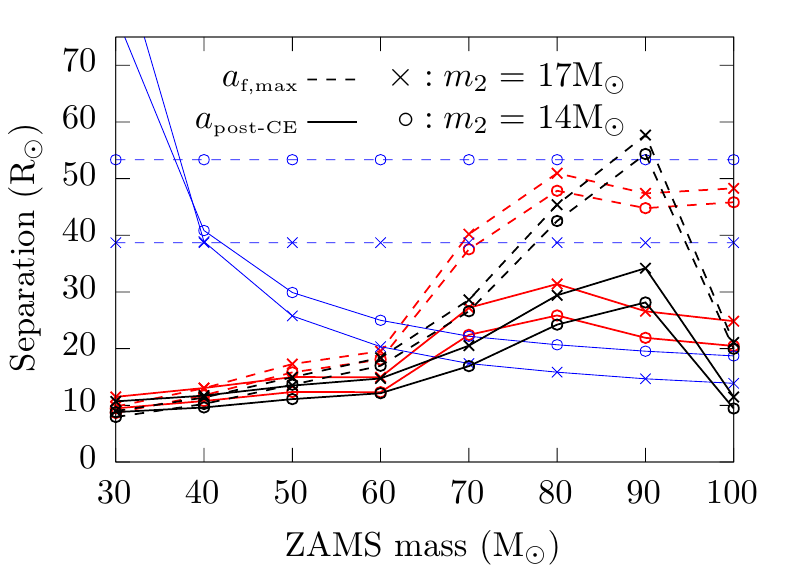}
 \caption{Post-CE (solid) and the largest possible final separation (dashed) of the binary as a function of the ZAMS mass. Colours indicate the applied wind scheme using the same colours as in Figure \ref{fig:remmass}. Separations are calculated using Equations (\ref{eq:sep}) and (\ref{eq:finsep}) with $m_2=14\msun$ (circles) and $m_2=17\msun$ (crosses). Blue lines show the separation required for the XRB phase calculated from Equations (\ref{eq:inisep}) and (\ref{eq:eggleton}), with $m_\textrm{BH}=m_\textrm{c}$ (solid) and $m_\textrm{BH}=0.8m_2$ (dashed).}\label{fig:a_f}
\end{figure}

In Figure \ref{fig:a_f} we show the post-CE separation and the upper limit of the final separation for each of our \texttt{MESA} models. It can be seen that the post-CE separation increases with mass due to the increase in core mass, except for the most massive cases ($\gtrsim80\msun$) where the binding energy increases more rapidly. The true final separation should be somewhere between the solid and dashed lines, presumably closer to the dashed lines if we believe the BH masses obtained in Figure \ref{fig:remmass}. By comparing these separations with the required separations shown in blue, it can be seen that the models in the range $\gtrsim70\msun$ satisfy the requirements. Note that the blue lines are just reference models assuming fully non-conservative MT for the XRB phase. The required separations can increase slightly depending on the BH mass retention rate.

From the above discussions, we have found a possible evolutionary path towards  a system with a large BH component. The binary likely originated from a binary with two massive stars and experienced CE evolution. The primary star should have been $\gtrsim70\msun$ to leave a large enough BH mass and end with a wide enough separation for the XRB phase. Stars in this mass range compose only $\lesssim0.1\%$ of the whole stellar population assuming a Salpeter-like initial mass function, which indicates that iPTF13bvn might have been a rare case.

\section{Conclusion}\label{sec:conclusion}
We have discussed the evolution of the binary that led to the supernova iPTF13bvn, based on the scenario proposed in \citet{RH17}. According to this scenario, the SN was caused by the initially less massive star in the binary, and the more massive star has already collapsed to a BH. We give constraints to the binary parameters for the second stage of the evolution, which should have been like an X-ray binary. The orbital separation at the beginning of this XRB phase should be $\gtrsim17$--$50\rsun$ depending on the mass of the BH. The BH mass is also unknown, but we estimate that it should have been relatively large to ensure stable mass transfer in the XRB phase.

We also found possible evolutionary paths to create binaries satisfying the above constraints. In order to create a large enough BH for this XRB phase, the initially more massive star should have been $\gtrsim70\msun$. The system most likely experienced a CE phase, and the mass range in which the post-CE separation meets the requirements for the XRB phase is yet agian $\gtrsim70\msun$. The number of binaries with such large masses are expected to be very small ($\lesssim0.1\%$). Thus if the progenitor of iPTF13bvn was truly formed through this path, it was a peculiar case and normal type Ib SNe should have been created through different pathways. This may provide a reason for the rarity of progenitor detection for type Ib SNe. Deeper observational constraints on the remaining companion are needed to determine the true evolutionary scenario for the progenitor of iPTF13bvn.

\section*{Acknowledgements}
The author thanks the Yukawa Institute for Theoretical Physics at Kyoto University. Discussions during the YITP workshop YITP-T-16-05 on "Transient Universe in
the Big Survey Era" were useful to complete this work. The author was supported by the JSPS Research fellowship for young scientists (DC2, No 16J07613).

\bibliographystyle{mnras}

\begin{thebibliography}{}
\makeatletter
\relax
\def\mn@urlcharsother{\let\do\@makeother \do\$\do\&\do\#\do\^\do\_\do\%\do\~}
\def\mn@doi{\begingroup\mn@urlcharsother \@ifnextchar [ {\mn@doi@}
  {\mn@doi@[]}}
\def\mn@doi@[#1]#2{\def\@tempa{#1}\ifx\@tempa\@empty \href
  {http://dx.doi.org/#2} {doi:#2}\else \href {http://dx.doi.org/#2} {#1}\fi
  \endgroup}
\def\mn@eprint#1#2{\mn@eprint@#1:#2::\@nil}
\def\mn@eprint@arXiv#1{\href {http://arxiv.org/abs/#1} {{\tt arXiv:#1}}}
\def\mn@eprint@dblp#1{\href {http://dblp.uni-trier.de/rec/bibtex/#1.xml}
  {dblp:#1}}
\def\mn@eprint@#1:#2:#3:#4\@nil{\def\@tempa {#1}\def\@tempb {#2}\def\@tempc
  {#3}\ifx \@tempc \@empty \let \@tempc \@tempb \let \@tempb \@tempa \fi \ifx
  \@tempb \@empty \def\@tempb {arXiv}\fi \@ifundefined
  {mn@eprint@\@tempb}{\@tempb:\@tempc}{\expandafter \expandafter \csname
  mn@eprint@\@tempb\endcsname \expandafter{\@tempc}}}

\bibitem[\protect\citeauthoryear{{Adams}, {Kochanek}, {Gerke}, {Stanek}  \&
  {Dai}}{{Adams} et~al.}{2016a}]{ada16a}
{Adams} S.~M.,  {Kochanek} C.~S.,  {Gerke} J.~R.,  {Stanek} K.~Z.,   {Dai} X.,
  2016a, preprint, \href {http://adsabs.harvard.edu/abs/2016arXiv160901283A} {}
  (\mn@eprint {arXiv} {1609.01283})

\bibitem[\protect\citeauthoryear{{Adams}, {Kochanek}, {Gerke}  \&
  {Stanek}}{{Adams} et~al.}{2016b}]{ada16b}
{Adams} S.~M.,  {Kochanek} C.~S.,  {Gerke} J.~R.,   {Stanek} K.~Z.,  2016b,
  preprint, \href {http://adsabs.harvard.edu/abs/2016arXiv161002402A} {}
  (\mn@eprint {arXiv} {1610.02402})

\bibitem[\protect\citeauthoryear{{Beasor} \& {Davies}}{{Beasor} \&
  {Davies}}{2016}]{bea16}
{Beasor} E.~R.,  {Davies} B.,  2016, \mn@doi [\mnras] {10.1093/mnras/stw2054},
  \href {http://adsabs.harvard.edu/abs/2016MNRAS.463.1269B} {463, 1269}

\bibitem[\protect\citeauthoryear{{Benvenuto}, {Bersten}  \&
  {Nomoto}}{{Benvenuto} et~al.}{2013}]{ben13}
{Benvenuto} O.~G.,  {Bersten} M.~C.,   {Nomoto} K.,  2013, \mn@doi [\apj]
  {10.1088/0004-637X/762/2/74}, \href
  {http://adsabs.harvard.edu/abs/2013ApJ...762...74B} {762, 74}

\bibitem[\protect\citeauthoryear{{Bersten} et~al.,}{{Bersten}
  et~al.}{2014}]{ber14}
{Bersten} M.~C.,  et~al., 2014, \mn@doi [\aj] {10.1088/0004-6256/148/4/68},
  \href {http://adsabs.harvard.edu/abs/2014AJ....148...68B} {148, 68}

\bibitem[\protect\citeauthoryear{{Cao} et~al.,}{{Cao} et~al.}{2013}]{cao13}
{Cao} Y.,  et~al., 2013, \mn@doi [\apjl] {10.1088/2041-8205/775/1/L7}, \href
  {http://adsabs.harvard.edu/abs/2013ApJ...775L...7C} {775, L7}

\bibitem[\protect\citeauthoryear{{Chini}, {Hoffmeister}, {Nasseri}, {Stahl}  \&
  {Zinnecker}}{{Chini} et~al.}{2012}]{chi12}
{Chini} R.,  {Hoffmeister} V.~H.,  {Nasseri} A.,  {Stahl} O.,   {Zinnecker} H.,
   2012, \mn@doi [\mnras] {10.1111/j.1365-2966.2012.21317.x}, \href
  {http://adsabs.harvard.edu/abs/2012MNRAS.424.1925C} {424, 1925}

\bibitem[\protect\citeauthoryear{{De Vito} \& {Benvenuto}}{{De Vito} \&
  {Benvenuto}}{2012}]{dev12}
{De Vito} M.~A.,  {Benvenuto} O.~G.,  2012, \mn@doi [\mnras]
  {10.1111/j.1365-2966.2012.20459.x}, \href
  {http://adsabs.harvard.edu/abs/2012MNRAS.421.2206D} {421, 2206}

\bibitem[\protect\citeauthoryear{{Eggleton}}{{Eggleton}}{1983}]{egg83}
{Eggleton} P.~P.,  1983, \mn@doi [\apj] {10.1086/160960}, \href
  {http://adsabs.harvard.edu/abs/1983ApJ...268..368E} {268, 368}

\bibitem[\protect\citeauthoryear{{Eldridge} \& {Maund}}{{Eldridge} \&
  {Maund}}{2016}]{eld16}
{Eldridge} J.~J.,  {Maund} J.~R.,  2016, \mn@doi [\mnras]
  {10.1093/mnrasl/slw099}, \href
  {http://adsabs.harvard.edu/abs/2016MNRAS.461L.117E} {461, L117}

\bibitem[\protect\citeauthoryear{{Eldridge}, {Fraser}, {Maund}  \&
  {Smartt}}{{Eldridge} et~al.}{2015}]{eld15}
{Eldridge} J.~J.,  {Fraser} M.,  {Maund} J.~R.,   {Smartt} S.~J.,  2015,
  \mn@doi [\mnras] {10.1093/mnras/stu2197}, \href
  {http://adsabs.harvard.edu/abs/2015MNRAS.446.2689E} {446, 2689}

\bibitem[\protect\citeauthoryear{{Ertl}, {Janka}, {Woosley}, {Sukhbold}  \&
  {Ugliano}}{{Ertl} et~al.}{2016}]{ert16}
{Ertl} T.,  {Janka} H.-T.,  {Woosley} S.~E.,  {Sukhbold} T.,   {Ugliano} M.,
  2016, \mn@doi [\apj] {10.3847/0004-637X/818/2/124}, \href
  {http://adsabs.harvard.edu/abs/2016ApJ...818..124E} {818, 124}

\bibitem[\protect\citeauthoryear{{Farmer}, {Fields}, {Petermann}, {Dessart},
  {Cantiello}, {Paxton}  \& {Timmes}}{{Farmer} et~al.}{2016}]{far16}
{Farmer} R.,  {Fields} C.~E.,  {Petermann} I.,  {Dessart} L.,  {Cantiello} M.,
  {Paxton} B.,   {Timmes} F.~X.,  2016, \mn@doi [\apjs]
  {10.3847/1538-4365/227/2/22}, \href
  {http://adsabs.harvard.edu/abs/2016ApJS..227...22F} {227, 22}

\bibitem[\protect\citeauthoryear{{Folatelli} et~al.,}{{Folatelli}
  et~al.}{2016}]{fol16}
{Folatelli} G.,  et~al., 2016, \mn@doi [\apjl] {10.3847/2041-8205/825/2/L22},
  \href {http://adsabs.harvard.edu/abs/2016ApJ...825L..22F} {825, L22}

\bibitem[\protect\citeauthoryear{{Fremling} et~al.,}{{Fremling}
  et~al.}{2014}]{fre14}
{Fremling} C.,  et~al., 2014, \mn@doi [\aap] {10.1051/0004-6361/201423884},
  \href {http://adsabs.harvard.edu/abs/2014A%26A...565A.114F} {565, A114}

\bibitem[\protect\citeauthoryear{{Fremling} et~al.,}{{Fremling}
  et~al.}{2016}]{fre16}
{Fremling} C.,  et~al., 2016, \mn@doi [\aap] {10.1051/0004-6361/201628275},
  \href {http://adsabs.harvard.edu/abs/2016A%26A...593A..68F} {593, A68}

\bibitem[\protect\citeauthoryear{{Fryer}}{{Fryer}}{1999}]{fry99}
{Fryer} C.~L.,  1999, \mn@doi [\apj] {10.1086/307647}, \href
  {http://adsabs.harvard.edu/abs/1999ApJ...522..413F} {522, 413}

\bibitem[\protect\citeauthoryear{{Fryer}, {Belczynski}, {Wiktorowicz},
  {Dominik}, {Kalogera}  \& {Holz}}{{Fryer} et~al.}{2012}]{fry12}
{Fryer} C.~L.,  {Belczynski} K.,  {Wiktorowicz} G.,  {Dominik} M.,  {Kalogera}
  V.,   {Holz} D.~E.,  2012, \mn@doi [\apj] {10.1088/0004-637X/749/1/91}, \href
  {http://adsabs.harvard.edu/abs/2012ApJ...749...91F} {749, 91}

\bibitem[\protect\citeauthoryear{{Ge}, {Hjellming}, {Webbink}, {Chen}  \&
  {Han}}{{Ge} et~al.}{2010}]{ge10}
{Ge} H.,  {Hjellming} M.~S.,  {Webbink} R.~F.,  {Chen} X.,   {Han} Z.,  2010,
  \mn@doi [\apj] {10.1088/0004-637X/717/2/724}, \href
  {http://adsabs.harvard.edu/abs/2010ApJ...717..724G} {717, 724}

\bibitem[\protect\citeauthoryear{{Groh}, {Georgy}  \& {Ekstr{\"o}m}}{{Groh}
  et~al.}{2013}]{gro13}
{Groh} J.~H.,  {Georgy} C.,   {Ekstr{\"o}m} S.,  2013, \mn@doi [\aap]
  {10.1051/0004-6361/201322369}, \href
  {http://adsabs.harvard.edu/abs/2013A%26A...558L...1G} {558, L1}

\bibitem[\protect\citeauthoryear{{Heger}, {Fryer}, {Woosley}, {Langer}  \&
  {Hartmann}}{{Heger} et~al.}{2003}]{heg03}
{Heger} A.,  {Fryer} C.~L.,  {Woosley} S.~E.,  {Langer} N.,   {Hartmann} D.~H.,
   2003, \mn@doi [\apj] {10.1086/375341}, \href
  {http://adsabs.harvard.edu/abs/2003ApJ...591..288H} {591, 288}

\bibitem[\protect\citeauthoryear{{Hirai}}{{Hirai}}{2017}]{RH17}
{Hirai} R.,  2017, \mn@doi [\mnras] {10.1093/mnras/stw3321}, \href
  {http://adsabs.harvard.edu/abs/2017MNRAS.466.3775H} {466, 3775}

\bibitem[\protect\citeauthoryear{{Hirai} \& {Yamada}}{{Hirai} \&
  {Yamada}}{2015}]{RH15}
{Hirai} R.,  {Yamada} S.,  2015, \mn@doi [\apj] {10.1088/0004-637X/805/2/170},
  \href {http://adsabs.harvard.edu/abs/2015ApJ...805..170H} {805, 170}

\bibitem[\protect\citeauthoryear{{Hjellming}}{{Hjellming}}{1989}]{hje89}
{Hjellming} M.~S.,  1989, PhD thesis, Illinois Univ.~at Urbana-Champaign,
  Savoy.

\bibitem[\protect\citeauthoryear{{Horiuchi}, {Nakamura}, {Takiwaki}, {Kotake}
  \& {Tanaka}}{{Horiuchi} et~al.}{2014}]{hor14}
{Horiuchi} S.,  {Nakamura} K.,  {Takiwaki} T.,  {Kotake} K.,   {Tanaka} M.,
  2014, \mn@doi [\mnras] {10.1093/mnrasl/slu146}, \href
  {http://adsabs.harvard.edu/abs/2014MNRAS.445L..99H} {445, L99}

\bibitem[\protect\citeauthoryear{{Iben} \& {Tutukov}}{{Iben} \&
  {Tutukov}}{1984}]{ibe84}
{Iben} Jr. I.,  {Tutukov} A.~V.,  1984, \mn@doi [\apj] {10.1086/162455}, \href
  {http://adsabs.harvard.edu/abs/1984ApJ...284..719I} {284, 719}

\bibitem[\protect\citeauthoryear{{Ivanova}}{{Ivanova}}{2011}]{iva11}
{Ivanova} N.,  2011, \mn@doi [\apj] {10.1088/0004-637X/730/2/76}, \href
  {http://adsabs.harvard.edu/abs/2011ApJ...730...76I} {730, 76}

\bibitem[\protect\citeauthoryear{{Ivanova} \& {Taam}}{{Ivanova} \&
  {Taam}}{2004}]{iva04}
{Ivanova} N.,  {Taam} R.~E.,  2004, \mn@doi [\apj] {10.1086/380561}, \href
  {http://adsabs.harvard.edu/abs/2004ApJ...601.1058I} {601, 1058}

\bibitem[\protect\citeauthoryear{{Ivanova} et~al.,}{{Ivanova}
  et~al.}{2013}]{iva13}
{Ivanova} N.,  et~al., 2013, \mn@doi [\aapr] {10.1007/s00159-013-0059-2}, \href
  {http://adsabs.harvard.edu/abs/2013A%26ARv..21...59I} {21, 59}

\bibitem[\protect\citeauthoryear{{Jiang}, {Stone}  \& {Davis}}{{Jiang}
  et~al.}{2014}]{jia14}
{Jiang} Y.-F.,  {Stone} J.~M.,   {Davis} S.~W.,  2014, \mn@doi [\apj]
  {10.1088/0004-637X/796/2/106}, \href
  {http://adsabs.harvard.edu/abs/2014ApJ...796..106J} {796, 106}

\bibitem[\protect\citeauthoryear{{Kochanek}}{{Kochanek}}{2014}]{koc14}
{Kochanek} C.~S.,  2014, \mn@doi [\apj] {10.1088/0004-637X/785/1/28}, \href
  {http://adsabs.harvard.edu/abs/2014ApJ...785...28K} {785, 28}

\bibitem[\protect\citeauthoryear{{Kochanek}}{{Kochanek}}{2015}]{koc15}
{Kochanek} C.~S.,  2015, \mn@doi [\mnras] {10.1093/mnras/stu2056}, \href
  {http://adsabs.harvard.edu/abs/2015MNRAS.446.1213K} {446, 1213}

\bibitem[\protect\citeauthoryear{{Kudritzki}, {Pauldrach}, {Puls}  \&
  {Abbott}}{{Kudritzki} et~al.}{1989}]{kud89}
{Kudritzki} R.~P.,  {Pauldrach} A.,  {Puls} J.,   {Abbott} D.~C.,  1989, \aap,
  \href {http://adsabs.harvard.edu/abs/1989A%26A...219..205K} {219, 205}

\bibitem[\protect\citeauthoryear{{Kuncarayakti} et~al.,}{{Kuncarayakti}
  et~al.}{2015}]{kun15}
{Kuncarayakti} H.,  et~al., 2015, \mn@doi [\aap] {10.1051/0004-6361/201425604},
  \href {http://adsabs.harvard.edu/abs/2015A%26A...579A..95K} {579, A95}

\bibitem[\protect\citeauthoryear{{Lovegrove} \& {Woosley}}{{Lovegrove} \&
  {Woosley}}{2013}]{lov13}
{Lovegrove} E.,  {Woosley} S.~E.,  2013, \mn@doi [\apj]
  {10.1088/0004-637X/769/2/109}, \href
  {http://adsabs.harvard.edu/abs/2013ApJ...769..109L} {769, 109}

\bibitem[\protect\citeauthoryear{{Lyman}, {Bersier}, {James}, {Mazzali},
  {Eldridge}, {Fraser}  \& {Pian}}{{Lyman} et~al.}{2016}]{lym16}
{Lyman} J.~D.,  {Bersier} D.,  {James} P.~A.,  {Mazzali} P.~A.,  {Eldridge}
  J.~J.,  {Fraser} M.,   {Pian} E.,  2016, \mn@doi [\mnras]
  {10.1093/mnras/stv2983}, \href
  {http://adsabs.harvard.edu/abs/2016MNRAS.457..328L} {457, 328}

\bibitem[\protect\citeauthoryear{{Nadezhin}}{{Nadezhin}}{1980}]{nad80}
{Nadezhin} D.~K.,  1980, \mn@doi [\apss] {10.1007/BF00638971}, \href
  {http://adsabs.harvard.edu/abs/1980Ap%26SS..69..115N} {69, 115}

\bibitem[\protect\citeauthoryear{{Nugis} \& {Lamers}}{{Nugis} \&
  {Lamers}}{2000}]{nug00}
{Nugis} T.,  {Lamers} H.~J.~G.~L.~M.,  2000, \aap, \href
  {http://adsabs.harvard.edu/abs/2000A%26A...360..227N} {360, 227}

\bibitem[\protect\citeauthoryear{{O'Connor} \& {Ott}}{{O'Connor} \&
  {Ott}}{2011}]{oco11}
{O'Connor} E.,  {Ott} C.~D.,  2011, \mn@doi [\apj]
  {10.1088/0004-637X/730/2/70}, \href
  {http://adsabs.harvard.edu/abs/2011ApJ...730...70O} {730, 70}

\bibitem[\protect\citeauthoryear{{Ohsuga}, {Mori}, {Nakamoto}  \&
  {Mineshige}}{{Ohsuga} et~al.}{2005}]{ohs05}
{Ohsuga} K.,  {Mori} M.,  {Nakamoto} T.,   {Mineshige} S.,  2005, \mn@doi
  [\apj] {10.1086/430728}, \href
  {http://adsabs.harvard.edu/abs/2005ApJ...628..368O} {628, 368}

\bibitem[\protect\citeauthoryear{{Papish}, {Soker}  \& {Bukay}}{{Papish}
  et~al.}{2015}]{pap15}
{Papish} O.,  {Soker} N.,   {Bukay} I.,  2015, \mn@doi [\mnras]
  {10.1093/mnras/stv345}, \href
  {http://adsabs.harvard.edu/abs/2015MNRAS.449..288P} {449, 288}

\bibitem[\protect\citeauthoryear{{Paxton}, {Bildsten}, {Dotter}, {Herwig},
  {Lesaffre}  \& {Timmes}}{{Paxton} et~al.}{2011}]{MESA1}
{Paxton} B.,  {Bildsten} L.,  {Dotter} A.,  {Herwig} F.,  {Lesaffre} P.,
  {Timmes} F.,  2011, \mn@doi [\apjs] {10.1088/0067-0049/192/1/3}, \href
  {http://adsabs.harvard.edu/abs/2011ApJS..192....3P} {192, 3}

\bibitem[\protect\citeauthoryear{{Paxton} et~al.,}{{Paxton}
  et~al.}{2013}]{MESA2}
{Paxton} B.,  et~al., 2013, \mn@doi [\apjs] {10.1088/0067-0049/208/1/4}, \href
  {http://adsabs.harvard.edu/abs/2013ApJS..208....4P} {208, 4}

\bibitem[\protect\citeauthoryear{{Paxton} et~al.,}{{Paxton}
  et~al.}{2015}]{MESA3}
{Paxton} B.,  et~al., 2015, \mn@doi [\apjs] {10.1088/0067-0049/220/1/15}, \href
  {http://adsabs.harvard.edu/abs/2015ApJS..220...15P} {220, 15}

\bibitem[\protect\citeauthoryear{{Postnov} \& {Yungelson}}{{Postnov} \&
  {Yungelson}}{2014}]{pos14}
{Postnov} K.~A.,  {Yungelson} L.~R.,  2014, \mn@doi [Living Reviews in
  Relativity] {10.12942/lrr-2014-3}, \href
  {http://adsabs.harvard.edu/abs/2014LRR....17....3P} {17, 3}

\bibitem[\protect\citeauthoryear{{Sana} et~al.,}{{Sana} et~al.}{2014}]{san14}
{Sana} H.,  et~al., 2014, \mn@doi [\apjs] {10.1088/0067-0049/215/1/15}, \href
  {http://adsabs.harvard.edu/abs/2014ApJS..215...15S} {215, 15}

\bibitem[\protect\citeauthoryear{{S{\c a}dowski} \& {Narayan}}{{S{\c a}dowski}
  \& {Narayan}}{2016}]{sad16}
{S{\c a}dowski} A.,  {Narayan} R.,  2016, \mn@doi [\mnras]
  {10.1093/mnras/stv2941}, \href
  {http://adsabs.harvard.edu/abs/2016MNRAS.456.3929S} {456, 3929}

\bibitem[\protect\citeauthoryear{{Smartt}, {Eldridge}, {Crockett}  \&
  {Maund}}{{Smartt} et~al.}{2009}]{sma09}
{Smartt} S.~J.,  {Eldridge} J.~J.,  {Crockett} R.~M.,   {Maund} J.~R.,  2009,
  \mn@doi [\mnras] {10.1111/j.1365-2966.2009.14506.x}, \href
  {http://adsabs.harvard.edu/abs/2009MNRAS.395.1409S} {395, 1409}

\bibitem[\protect\citeauthoryear{{Soker}}{{Soker}}{2015}]{sok15}
{Soker} N.,  2015, \mn@doi [\apj] {10.1088/0004-637X/800/2/114}, \href
  {http://adsabs.harvard.edu/abs/2015ApJ...800..114S} {800, 114}

\bibitem[\protect\citeauthoryear{{Srivastav}, {Anupama}  \& {Sahu}}{{Srivastav}
  et~al.}{2014}]{sri14}
{Srivastav} S.,  {Anupama} G.~C.,   {Sahu} D.~K.,  2014, \mn@doi [\mnras]
  {10.1093/mnras/stu1878}, \href
  {http://adsabs.harvard.edu/abs/2014MNRAS.445.1932S} {445, 1932}

\bibitem[\protect\citeauthoryear{{Sukhbold}, {Ertl}, {Woosley}, {Brown}  \&
  {Janka}}{{Sukhbold} et~al.}{2016}]{suk16}
{Sukhbold} T.,  {Ertl} T.,  {Woosley} S.~E.,  {Brown} J.~M.,   {Janka} H.-T.,
  2016, \mn@doi [\apj] {10.3847/0004-637X/821/1/38}, \href
  {http://adsabs.harvard.edu/abs/2016ApJ...821...38S} {821, 38}

\bibitem[\protect\citeauthoryear{{Ugliano}, {Janka}, {Marek}  \&
  {Arcones}}{{Ugliano} et~al.}{2012}]{ugl12}
{Ugliano} M.,  {Janka} H.-T.,  {Marek} A.,   {Arcones} A.,  2012, \mn@doi
  [\apj] {10.1088/0004-637X/757/1/69}, \href
  {http://adsabs.harvard.edu/abs/2012ApJ...757...69U} {757, 69}

\bibitem[\protect\citeauthoryear{{Vink}, {de Koter}  \& {Lamers}}{{Vink}
  et~al.}{2001}]{vin01}
{Vink} J.~S.,  {de Koter} A.,   {Lamers} H.~J.~G.~L.~M.,  2001, \mn@doi [\aap]
  {10.1051/0004-6361:20010127}, \href
  {http://adsabs.harvard.edu/abs/2001A%26A...369..574V} {369, 574}

\bibitem[\protect\citeauthoryear{{Webbink}}{{Webbink}}{1984}]{web84}
{Webbink} R.~F.,  1984, \mn@doi [\apj] {10.1086/161701}, \href
  {http://adsabs.harvard.edu/abs/1984ApJ...277..355W} {277, 355}

\bibitem[\protect\citeauthoryear{{Webbink}}{{Webbink}}{1985}]{web85}
{Webbink} R.~F.,  1985, {Stellar evolution and binaries}.
p.~39

\makeatother
\end{thebibliography}

\bsp
\label{lastpage}
\end{document}